\documentclass[conference]{IEEEtran}
\IEEEoverridecommandlockouts
\usepackage{cite}
\usepackage{amsmath,amssymb,amsfonts}
\usepackage{algorithm}
\usepackage{algorithmic}

\usepackage{graphicx}
\usepackage{textcomp}
\usepackage{xcolor}
\usepackage{url}
\usepackage{booktabs}
\usepackage{multirow}
\def\BibTeX{{\rm B\kern-.05em{\sc i\kern-.025em b}\kern-.08em
    T\kern-.1667em\lower.7ex\hbox{E}\kern-.125emX}}
\begin{document}

\title{PILOT: One Physics-Integrated Generation Framework to Unify 2D and 3D Radio Map Construction}

\author{
\IEEEauthorblockN{
Weiming Huang\IEEEauthorrefmark{1},
Hao Sun\IEEEauthorrefmark{2},
Junting Chen\IEEEauthorrefmark{1}
}
\IEEEauthorblockA{\IEEEauthorrefmark{1}
School of Science and Engineering (SSE) and Shenzhen Future Network of Intelligence Institute (FNii-Shenzhen)\\
The Chinese University of Hong Kong, Shenzhen, Guangdong 518172, China
}
\IEEEauthorblockA{\IEEEauthorrefmark{2}
Department of Electrical Engineering, City University of Hong Kong, Hong Kong
}
}

\maketitle

\begin{abstract}
Unified 2D and 3D radio map construction supports network planning, wireless
digital twins, and unmanned aerial vehicle (UAV) applications. In urban
environments, blockage, reflection, and diffraction make accurate construction
expensive for physics-based solvers. Autoregressive next-token prediction offers
a single sequential formulation that can cover both 2D and 3D generation, but
standard raster ordering ignores the spatial structure of radio propagation.
When generation follows propagation, each token is predicted from
propagation-relevant history rather than spatially arbitrary context, which
provides more causally informative conditioning and lowers conditional
uncertainty. We propose PILOT, a pretrained autoregressive framework that
replaces raster scan with a wavefront sequence expanding outward from the
transmitter. Each prediction step is guided by an environment-aware instruction
that spatially aligns environment features with the queried radio map region.
The same framework extends to 3D radio maps through height-slice stacking while
a gradient loss enforces vertical continuity. On standard 2D benchmarks, PILOT
achieves the lowest NMSE among all baselines. For volumetric generation, it
reduces NMSE by 78\% relative to the diffusion baseline at roughly $2500\times$
faster inference. It also outperforms methods that rely on 10\% sparse
measurements and achieves the best zero-shot results in the cross-domain
evaluation. \noindent\textit{Code:} \url{https://github.com/Uminan/PILOT}
\end{abstract}
\begin{IEEEkeywords}
radio map construction, pathloss estimation, autoregressive generation, wireless communications
\end{IEEEkeywords}

\section{Introduction}
As sixth-generation wireless networks extend coverage from the ground plane to
aerial and multi-floor users, channel characterization needs to cover
3D regions rather than street-level slices
\cite{zhang2024machine,li2024fast}. Wireless digital twins, cellular-connected
unmanned aerial vehicle (UAV) networking, and volumetric coverage analysis
therefore require radio maps in both 2D and 3D over the same scene
\cite{wang2025digital,li2024fast}. Dense measurements over such regions are
infeasible: ground surveys are labor-intensive, and upper-floor interiors and
aerial corridors cannot be sampled as densely \cite{li2024fast}. By contrast,
environment maps encoding building geometry are available before deployment from
urban geographic databases and digital twin assets, and extend from planar to
volumetric scenes \cite{lee2024scalable}. In dense urban scenes, pathloss is
shaped by blockage, reflection, and diffraction around irregular structures, so
local pathloss depends on geometry well beyond the line-of-sight path. Classical
ray tracing models these interactions, but its cost grows rapidly with
resolution and multi-bounce order, becoming impractical for 3D volumes
\cite{lee2024scalable}.

In this paper, we focus on the construction of pathloss radio maps from scene
geometry and transmitter configuration, using one formulation for both 2D slices
and 3D volumes over the same urban scene. The map characterizes pathloss across
ground-level locations and low-altitude airspace, with the 2D slice and the 3D
field coinciding at shared coordinates. Wireless digital twins rely on such a
map for network planning, where one volumetric field must cover street-level
corridors and upper-floor interiors within a single urban block.
Cellular-connected unmanned aerial vehicle (UAV) networking relies on it for aerial base-station placement
and trajectory design, which depend on altitude-dependent pathloss rather than a
single ground-level slice. These applications demand efficient re-evaluation
when the carrier frequency or link-budget setting changes, together with
source-aware prediction in which the pathloss field over the scene is structured
by signal propagation from the transmitter.

For this geometry-to-pathloss construction setting, prior work mainly follows
physics-based simulation and learning-based prediction. Classical ray tracing
is the representative physics-based approach, modeling reflection, diffraction,
and scattering over a detailed 3D environment with encoded geometry and material
properties \cite{wu2025ckmimagenet}. Its fidelity, however, incurs a
computational cost that makes repeated 3D evaluation impractical
\cite{lee2024scalable}. Learning-based regressors reduce this cost by predicting
2D radio maps from scene geometry and transmitter location in a single forward
pass. RadioUNet employs a cascaded two-UNet architecture that maps city-map
geometry and transmitter location to the 2D radio map \cite{levie2021radiounet}.
RadioMamba retains this 2D regression formulation and replaces the convolutional neural network (CNN) backbone
with a hybrid Mamba-UNet to capture long-range spatial dependencies and to
improve the efficiency-accuracy trade-off \cite{jia2025radiomamba}. Other methods
instead reconstruct radio maps from sparse on-site measurements, with RME-GAN
using a two-phase conditional generative adversarial framework and deep
completion autoencoders inferring the field directly from measured samples
\cite{zhang2023rme,tegan2022deep}. Across these learning-based pipelines, construction remains
posed as direct regression or reconstruction rather than as a source-aware
process. Most methods are restricted to 2D prediction, and
measurement-conditioned variants further depend on on-site samples that are
difficult to acquire densely over the target region.

Among existing generative approaches, conditional diffusion is the
representative framework that formulates radio-map construction as iterative
denoising rather than direct regression or measurement-conditioned
reconstruction. RadioDiff encodes the radio map into a variational autoencoder (VAE) latent space and
applies a 2D UNet for denoising conditioned on environment geometry and
transmitter location. RadioDiff-3D extends the backbone to a 3D UNet for
volumetric generation. In both formulations, however, the denoising trajectory
follows a fixed noise schedule rather than a source-aware signal-propagation
process. Iterative denoising further grows costly as spatial resolution and
sampling steps increase, which conflicts with real-time construction
requirements.

\begin{figure}[htbp]
  \centering
  \includegraphics[width=0.85\columnwidth]{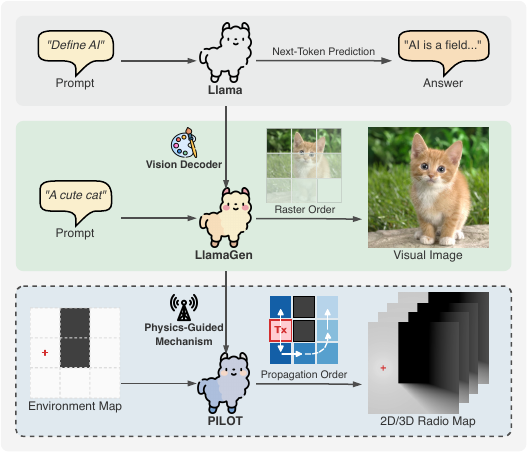}
  \caption{Overview of the PILOT framework.}
  \label{fig:PILOT_framework}
\end{figure}

This paper develops a physics-guided pretrained autoregressive framework for
source-aware radio map construction. Radio propagation guides the generation
process, so each token is predicted along propagation paths from the transmitter
rather than in raster order. Most existing methods use separate architectures
for different scenarios, each trained from scratch; an autoregressive decoder
consolidates these into one tokenized backbone that scales from 2D to 3D and
across carrier frequencies without redesign. By drawing on spatial priors from
large-scale visual pretraining, limited wireless supervision adapts the decoder
to propagation-induced effects, enabling zero-shot cross-domain transfer.
Fig.~\ref{fig:PILOT_framework} illustrates the progression from next-token
prediction in text and vision models to the proposed physics-integrated latent
ordered transformer (PILOT) framework. Recent LLM-based work has introduced
sequence modeling into radio-map construction, yet a propagation-aligned
mechanism is still not embedded in the generation process. LLM4PG \cite{sun2025llm4pg} uses an
LLM backbone as a single-pass regressor without an explicit generation order.
Ripple \cite{peng2025context} formulates radio map construction as in-context learning with a
frozen autoregressive vision model and a geometric spiral-out order, but relies
on same-layout demonstrations as prompts and ignores blockage, making it
unsuitable for the sampling-free cross-scene setting.

In visual autoregressive models, each token depends on spatially adjacent
context. In a radio field, however, the informative context for a target
location traces the propagation path from the transmitter and may bypass
neighboring regions entirely. This propagation dependency raises three
questions. How to guide the generation order so that the model follows the
physical propagation of radio signals? How to represent the radio field so that
it preserves both horizontal boundary fidelity and vertical continuity under
real-time constraints? How to keep prediction consistent when carrier frequency
and link-budget settings shift the coarse pathloss scale?

We tackle these challenges through a propagation-guided autoregressive framework
built on a pretrained visual backbone. To resolve the generation-order mismatch,
PILOT adopts a wavefront propagation order that prioritizes nearby low-blockage
regions before shadowed areas. This order reflects blockage-aware propagation
costs accumulated along paths from the transmitter to each target region, using
the building environment map. At each step, the radio token is predicted with
guidance from its spatially corresponding environment token, and spatial
coordinates are embedded via 3D rotary position embedding (3D-RoPE) to ensure positional registration and
semantic alignment. By adjusting the height coordinate in 3D-RoPE, we further
harness the framework to generate at the target height for various height-layer
applications. To address field structure complexity, a tokenizer compresses the
radio field into a discrete token space to accommodate both planar and
volumetric generation. Specifically, height information of the 3D map is encoded
along the channel dimension to avoid volumetric processing, assisted by a 3D
gradient regularizer that preserves boundary fidelity and vertical continuity.
To compensate for pathloss scale variation, PILOT inputs a pathloss anchor map
based on free-space path loss (FSPL) with shadowing that encodes carrier-frequency and pathloss-range
information to calibrate the coarse field scale. One architecture therefore
supports 2D and 3D construction and enables zero-shot transfer across carrier
frequencies and scene geometries without retraining.

To validate the role of generation order in training convergence and inference
accuracy, we construct a true-PL order that ranks tokens from low to high
ground-truth pathloss. This order serves as an oracle upper bound at inference,
yet training on it alone overfits. Mixing physics-guided orders improves
generalization over single-order training and further outperforms random
permutations and geometric orders such as raster and Z-curve orders. Among all
candidate orders, the wavefront order yields the lowest predictive entropy, with
entropy increasing monotonically over successive generation steps. This confirms
the design principle of wavefront order: prioritizing low-uncertainty tokens
first limits error accumulation. On the standard 2D benchmark, PILOT reduces
normalized mean square error (NMSE) by 9.5\% relative to RadioMamba \cite{jia2025radiomamba}. Under zero-shot transfer across carrier
frequency, pathloss range, and scene geometry, PILOT reduces NMSE by 7.9\%
relative to the closest zero-shot baseline, RME-GAN \cite{zhang2023rme}. For volumetric generation,
PILOT reduces NMSE by 78\% at roughly $2500\times$ faster inference than the
diffusion baseline \cite{wang2025radiodiff3d} and outperforms a variant with 10\% sparse measurements.

The main contributions are:
\begin{itemize}
    \item We propose PILOT, a pretrained autoregressive framework that unifies 2D and 3D radio map construction as environment-aware next-token prediction. To align radio and environment tokens in a shared spatial frame, physical coordinates are encoded into the autoregressive sequence through 3D-RoPE.

    \item We develop a wavefront propagation order that mirrors signal propagation from the transmitter along blockage-aware paths to guide the autoregressive generation sequence, so that each prediction step conditions on propagation-relevant context.

    \item We design a shared 2D tokenizer for both tasks by encoding height information in the channel dimension, thereby avoiding volumetric processing, while a 3D gradient regularizer preserves boundary fidelity and vertical continuity. The pathloss anchor map further calibrates the coarse field scale, thereby enabling zero-shot transfer across carrier frequencies and link-budget settings.

    \item On three ray-traced benchmarks, PILOT reduces NMSE by 9.5\% in standard 2D construction and by 7.9\% under zero-shot cross-domain transfer, relative to the respective state-of-the-art baselines. For 3D generation, it reduces NMSE by 78\% over the diffusion baseline while achieving approximately $2500\times$ faster inference in a sampling-free setting.
\end{itemize}

The rest of this paper is organized as follows. Section \ref{sec:formulation} presents the system
model and problem formulation. Section \ref{sec:pilot} details the proposed PILOT framework.
Section \ref{sec:evaluation} reports the experimental results and analysis. Section \ref{sec:conclusion} concludes
the paper. The core variables are summarized in Table \ref{tab:notation}.

\begin{table}[htbp]
\caption{Notation Table}
\label{tab:notation}
\centering
\resizebox{\columnwidth}{!}{%
\begin{tabular}{c|c}
\hline
Symbol & Description \\
\hline
$\mathbf{E}$ & Environment map \\
$\mathcal{H}$ & Building height map \\
$\mathcal{M}_{\mathrm{anc}}$ & Pathloss anchor map \\
$\mathbf{R},\;\hat{\mathbf{R}}$ & Ground-truth and predicted radio maps \\
$\mathbf{e}$ & Environment token \\
$\mathbf{r},\;\hat{\mathbf{r}}$ & Ground-truth and predicted radio tokens \\
$N_z$ & Number of height-slice channels \\
$\mathcal{F}_{\theta}$ & Autoregressive NN with parameters $\theta$ \\
$p_{\theta}$ & Autoregressive conditional distribution \\
$\pi$ & Generation-order permutation \\
$\mathcal{P}$ & Set of spatial patches \\
$\mathbf{u}$ & 3D spatial coordinate \\
$\beta(\cdot)$ & Blockage ratio along a ray \\
$\mathbb{I}(\cdot)$ & Indicator function \\
$D$ & Accumulated propagation cost \\
$\mathcal{N}$ & 8-connected neighborhood of a patch \\
$f$ & Carrier frequency \\
$\Delta_P$ & Pathloss range \\
$|\mathcal{C}|$ & Codebook size \\
$H(\cdot)$ & Predictive entropy function \\
$\bigoplus$ & Block-diagonal direct sum \\
$\lambda, \alpha$ & Hyperparameters \\
\hline
\end{tabular}%
}
\end{table}

\section{System Model and Problem Formulation}
\label{sec:formulation}
\subsection{Scenario and Task Definition}

We consider sampling-free radio map construction in dense urban environments
with static geometry and a single transmitter. Let $\mathbf{E}$ denote the
environment-side input consisting of the building map and transmitter
configuration. Let $\mathbf{R}, \hat{\mathbf{R}} \in \mathbb{R}^{S \times S
\times N_z}$ denote the ground-truth and predicted pathloss fields on an
$S \times S$ spatial grid. The field-level prediction task is
\begin{equation}
    \hat{\mathbf{R}}=\mathcal{F}_\theta(\mathbf{E}, z_{\mathrm{rx}}, N_z),
\end{equation}
where $\mathcal{F}_\theta$ denotes a neural network parameterized by $\theta$.
The model learns $\theta$ by minimizing the prediction loss
$\mathcal{L}(\hat{\mathbf{R}}, \mathbf{R})$.

This formulation unifies 2D and 3D radio map prediction in one field
representation. When $N_z=1$, the output is a 2D radio map at receiver height
$z_{\mathrm{rx}}$. When $N_z>1$, the output is a 3D radio map over the receiver-height range
determined by $z_{\mathrm{rx}}$ and $N_z$, where $z_{\mathrm{rx}}$ specifies
the center of the receiver-height range and $N_z$ determines the number of
predicted height levels.

\subsection{Physics-Aware Input}

The environment input is defined as
\begin{equation}
\mathbf{E}=[\mathcal{H},\,\mathcal{M}_{\mathrm{tx}},\,\mathcal{M}_{\mathrm{anc}}],
\end{equation}
where $\mathcal{H}$ is the building height map, $\mathcal{M}_{\mathrm{tx}}$ is
the transmitter position map, and $\mathcal{M}_{\mathrm{anc}}$ is the pathloss
anchor map. Prior methods are typically developed for a single domain with fixed
carrier frequency and link-budget setting, so the coarse pathloss scale is
absorbed by the training data distribution and need not be encoded explicitly in
the input. Under cross-domain shifts, the free-space attenuation varies with
carrier frequency $f$, while the valid pathloss range is further determined by
the link-budget setting of the target domain. The third channel
$\mathcal{M}_{\mathrm{anc}}$ is therefore introduced to provide a
frequency-aware anchor for the coarse pathloss scale.

The pathloss anchor map is defined as
\begin{equation}
\mathcal{M}_{\mathrm{anc}}(\mathbf{u};f,L_{\mathrm{thr}})
=L_{\mathrm{FSPL}}\!\left(\|\mathbf{u}-\mathbf{u}_{\mathrm{tx}}\|,f\right)
+L_{\mathrm{shd}}(\mathbf{u};f,L_{\mathrm{thr}}),
\end{equation}
where $L_{\mathrm{FSPL}}$ is the free-space pathloss and the shadow correction is
\begin{equation}
L_{\mathrm{shd}}(\mathbf{u};f,L_{\mathrm{thr}})
=\beta(\mathbf{u}_{\mathrm{tx}},\mathbf{u})\bigl[L_{\mathrm{FSPL}}(d_0,f)-L_{\mathrm{thr}}\bigr].
\end{equation}
Here $L_{\mathrm{thr}}=10\log_{10}(WN_0)+NF-(P_{\mathrm{Tx}})_{\mathrm{dB}}$ is
a domain-dependent threshold determined by the link-budget setting of the target
domain, where $W$ is the bandwidth, $NF$ is the noise figure, and
$P_{\mathrm{Tx}}$ is the transmit power; and $d_0$ is a near-field reference
distance. The correction $L_{\mathrm{shd}}$ uses the blockage ratio $\beta$ to
scale the shadow range $[L_{\mathrm{FSPL}}(d_0,f)-L_{\mathrm{thr}}]$, so that
$\mathcal{M}_{\mathrm{anc}}$ encodes the frequency-aware coarse pathloss scale
that $\mathcal{H}$ and $\mathcal{M}_{\mathrm{tx}}$ cannot convey.

\begin{figure}[t]
  \centering
  \includegraphics[width=\columnwidth]{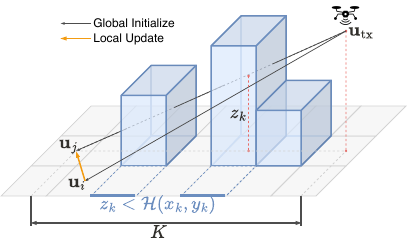}
  \caption{Blockage ratio computation. $K$ sample points are placed along the ground-plane projection from $\mathbf{u}_{\mathrm{tx}}$ to $\mathbf{u}_j$; blue segments indicate blocked portions where $z_k < \mathcal{H}(x_k,y_k)$.}
  \label{fig:blockage_ratio}
\end{figure}

The blockage ratio is defined as
\begin{equation}
\beta(\mathbf{u}_{\mathrm{tx}},\mathbf{u})
=\frac{1}{K}\sum_{k=1}^{K}\mathbb{I}\!\left(z_k<\mathcal{H}(x_k,y_k)\right),
\end{equation}
where $(x_k,y_k,z_k)$ is the $k$-th sample along the ray from
$\mathbf{u}_{\mathrm{tx}}$ to $\mathbf{u}$, $\mathcal{H}(x_k,y_k)$ is the
building height at that sample, $K$ is determined by the ray length and grid
resolution, and $\mathbb{I}(\cdot)$ is the indicator function. The
quantity $\beta(\mathbf{u}_{\mathrm{tx}},\mathbf{u})$ measures the blocked
fraction of the direct path and scales the shadow correction $L_{\mathrm{shd}}$,
as illustrated in Fig.~\ref{fig:blockage_ratio}. As a result, the anchor map
$\mathcal{M}_{\mathrm{anc}}$ varies with the blockage level along the direct
transmitter-to-target path.

\subsection{Order-Aware Sequence Formulation}

Prior work \cite{wang2024radiodiff} formulates radio map construction as
continuous field prediction conditioned on $\mathbf{E}$. The relevant
dependencies in this field are long-range and propagation-dependent, since
blockage, reflection, and diffraction couple distant regions of the map rather
than purely local neighborhoods. The task is reformulated as a sequential
generation problem, in which the conditional distribution
\begin{equation}
p_{\theta}(\mathbf{R}\!\mid\!\mathbf{E})
\end{equation}
is realized by sequential decoding.

A single-pass discriminative predictor does not provide such an explicit
generation process. A diffusion model is generative, but its trajectory follows
a denoising schedule rather than a spatial generation order over the radio map.
Decoder-only autoregressive next-token prediction provides an explicit
next-token interface for modeling
$p_{\theta}(\mathbf{R}\!\mid\!\mathbf{E})$, which is the formulation adopted in
this work.

To match the autoregressive generation paradigm, the continuous radio map is
split into patches and represented by $N$ discrete radio tokens
$\mathbf{r}=[r_1,\ldots,r_N]$. The environment input is encoded into an
aligned sequence $\mathbf{e}=[e_1,\ldots,e_N]$ with the same spatial indexing
as $\mathbf{r}$. In the proposed pipeline, the vector-quantization (VQ) stage
first establishes this discrete target space, so that 2D and 3D radio map
construction share a common token interface for autoregressive prediction.

Given $\mathbf{r}$ and $\mathbf{e}$, the conditional distribution factorizes as
\begin{equation}
\label{eq:ar_factorization}
p_{\theta}(\mathbf{r}\!\mid\!\mathbf{e})
=
\prod_{n=1}^{N}
p_{\theta}(r_n\!\mid\!\mathbf{e}, r_1,\ldots,r_{n-1}),
\end{equation}
which defines the prefix conditioning structure under the default ordering
$n=1,\ldots,N$ but leaves the mapping from token index to spatial position
unspecified.

This mapping is not fixed by decoder-only autoregressive generation itself.
The work in \cite{pang2025randar} treats the generation order as a design
choice without changing the decoder. Let $\pi \in \mathfrak{S}_N$ be a
permutation of the $N$ token indices, so that $\pi(n)$ is the index predicted
at step $n$. The order-aware factorization is
\begin{equation}
\label{eq:ar_factorization_pi}
p_{\theta}(\mathbf{r}\!\mid\!\mathbf{e};\pi)
=
\prod_{n=1}^{N}
p_{\theta}\!\left(
r_{\pi(n)} \!\mid\! \mathbf{e}, r_{\pi(1)},\ldots,r_{\pi(n-1)}
\right),
\end{equation}
where $\mathbf{e}$ provides global context at every step and $\pi$ determines
which radio-token prefix is available when $r_{\pi(n)}$ is predicted.

The available prefix shapes the conditional uncertainty at each step. For a
radio field, earlier tokens are informative only when they are
propagation-relevant to the current target, so that the prefix encodes the
blockage and detour structure along the propagation path. Raster or arbitrary
orders may expose prefixes that are nearby in space yet irrelevant to the
propagation state of the current target, thereby increasing the conditional
uncertainty in shadowed regions. The remaining design question is how to construct $\pi$ so that generation
follows the propagation process from the transmitter. In particular,
propagation-relevant regions should be predicted before the shadowed regions
that depend on them.

\section{PILOT Network Architecture}
\label{sec:pilot}
In this section, we develop PILOT, an environment-aware
autoregressive architecture for radio map construction. A vanilla
decoder-only autoregressive predicts each token from a raster prefix that is often
misaligned with the propagation state of the current target, and offers
no built-in way for the environment $\mathbf{E}$ to shape either the
generation order or the per-step prediction. To align the order-aware
factorization in \eqref{eq:ar_factorization_pi} with radio propagation,
we need to address the following technical challenges:
\begin{itemize}
    \item How to convert the continuous radio field $\mathbf{R}$ into
    discrete tokens that retain sharp shadow boundaries and preserve
    vertical continuity for 3D maps formed by channel-stacked height
    slices, so that 2D and 3D maps share a single token space?
    \item How to construct the generation order $\pi$ from the
    environment $\mathbf{E}$ so that the prefix $r_{\pi(<n)}$ at each
    step exposes propagation-relevant context for predicting
    $r_{\pi(n)}$, rather than spatially nearby but propagation-irrelevant
    tokens?
    \item How to inject the environment $\mathbf{E}$ into the decoder so
    that, under the order $\pi$, each radio-token prediction is
    conditioned on environment evidence at the same spatial coordinate,
    in both 2D and 3D?
\end{itemize}

\begin{figure*}[t]
  \centering
  \includegraphics[width=\textwidth]{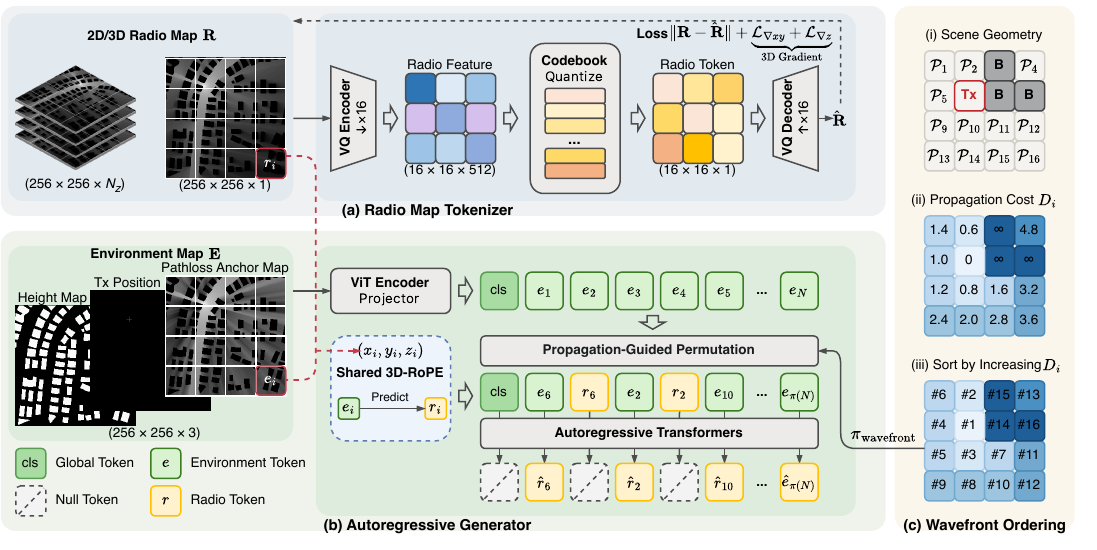}
  \caption{Overall framework of the proposed PILOT.}
  \label{fig:overall_framework}
\end{figure*}

To address these challenges, PILOT combines a blockage-aware graph
relaxation that constructs $\pi$ from $\mathbf{E}$, an autoregressive
decoder conditioned on environment evidence at every step, and a
tokenizer that yields a discrete space shared between 2D and 3D
maps. The overall architecture is shown in Fig.~\ref{fig:overall_framework}.

\subsection{Overall Pipeline}
\label{subsec:overall_pipeline}

PILOT decomposes radio map construction into a tokenizer and an
autoregressive generator. The tokenizer is trained first and then frozen,
after which the autoregressive generator is trained to predict indices
into the resulting fixed codebook. The tokenizer maps the continuous
radio map $\mathbf{R}\in\mathbb{R}^{256\times256\times N_z}$ to a
$16\times16$ grid of $N=256$ discrete tokens drawn from a shared
codebook, and reconstructs $\mathbf{R}$ from these tokens at inference.
Stacking $N_z$ height slices along the channel dimension allows the same
2D tokenizer to handle both 2D and 3D maps and produce $N=256$ radio
tokens in either case.

The wavefront ordering module takes the environment $\mathbf{E}$ and
outputs a permutation $\pi$ that ranks the $N$ patches by accumulated
propagation cost from the transmitter. The cost combines Euclidean
distance with a blockage penalty derived from $\mathcal{H}$ and
$\mathbf{u}_{\text{tx}}$, so $\pi$ visits propagation-relevant patches
before the shadowed patches that depend on them. At inference, this
order is fixed. During training, it is sampled together with other
candidate orders.

The autoregressive generator encodes $\mathbf{E}$ into a sequence of
environment tokens through a DINOv3-initialized vision transformer (ViT), interleaves them
with the radio tokens under the order $\pi$, and predicts the next
radio token at each step. A shared 3D rotary position embedding
registers each environment-radio token pair to its spatial coordinate,
so a single decoder handles both 2D and 3D maps. The frozen tokenizer
then decodes the predicted token sequence to $\hat{\mathbf{R}}$.

\subsection{Radio Map Tokenizer}
\label{subsec:radio_map_tokenizer}

The tokenizer compresses the radio map
$\mathbf{R}\in\mathbb{R}^{256\times256\times N_z}$ into a $16\times16$
latent grid through a 2D convolutional encoder, and reconstructs
$\hat{\mathbf{R}}$ from the latent grid through a 2D convolutional
decoder. Each spatial position of the latent grid is then quantized to
one entry of a shared codebook, yielding $N=256$ discrete tokens that
form the target space for the autoregressive stage and that are common
to 2D and 3D inputs. A 512-dimensional latent space is used to cover
the wide dynamic range of pathloss. We adopt the
LlamaGen~\cite{sun2024autoregressive} codebook size, $|C|=16384$.

However, raising the latent dimension under standard VQ-VAE training
causes severe codebook collapse, in which most entries remain unused.
PILOT introduces two design changes. First, the convolutional code
projector is replaced with a shared linear projector, which couples all
code vectors through the same trainable mapping and lets them evolve
jointly during training. Second, a differentiable vector quantization
scheme is adopted to restore gradient flow through the discrete
bottleneck without an auxiliary commitment loss. These changes mitigate
codebook collapse, as examined in Section~\ref{subsec:ablation_studies}.

For 3D inputs, channel-stacked height slices avoid volumetric
convolution and preserve a shared codebook with 2D maps, but they can
weaken the vertical smoothness of the reconstruction. To compensate,
the VQ-stage loss augments the reconstruction term with a multi-scale
gradient regularizer,
\begin{equation}
\label{eq:vq_loss}
\mathcal{L}_{\text{vq}}
=
\bigl\|\hat{\mathbf{R}}-\mathbf{R}\bigr\|
+
\lambda_{\text{grad}}\,\mathcal{L}_{\text{grad3D}},
\end{equation}
where the gradient term is
\begin{equation}
\label{eq:grad3d}
\begin{split}
  \mathcal{L}_{\mathrm{grad3D}}
  &= \sum_{s\in\mathcal{S}}
    \mathbb{E}\!\bigl[\|\nabla_{xy}\hat{\mathbf{R}}^{(s)}
    - \nabla_{xy}\mathbf{R}^{(s)}\|\bigr]\\
  &\quad+ \lambda_z\,
    \mathbb{E}\!\bigl[\|\nabla_z\hat{\mathbf{R}}
    - \nabla_z\mathbf{R}\|\bigr].
\end{split}
\end{equation}
Here, $\nabla_{xy}$ and $\nabla_z$ denote finite in-plane and vertical
differences, $\mathcal{S}$ is the set of downsampling scales used in
multi-scale evaluation, $\lambda_{\text{grad}}$ controls the weight of
the gradient term relative to reconstruction, and $\lambda_z$ sets the
relative importance of vertical versus in-plane gradients. For 2D maps
with $N_z=1$, the vertical term is dropped.

\subsection{Wavefront Propagation Order}
\label{subsec:wavefront_ordering}

Raster or fixed spatial orders do not reflect the physical nature of
radio propagation. The prefix available when predicting a shadowed
patch may then contain spatially nearby tokens that are weakly related
to its propagation state, while omitting the lower-cost regions that
physically shape it. PILOT therefore adopts a wavefront propagation
order, denoted by $\pi_{\text{wavefront}}$, which expands outward from
the transmitter along blockage-aware low-cost paths.

\subsubsection{Order Construction}
\label{subsubsec:wavefront_order_construction}

The wavefront order is defined by ranking patches according to their
accumulated blockage-aware propagation cost from the transmitter, so
that the sequence expands outward along lower-cost paths rather than
following a fixed spatial scan. At the initial step $t=0$, each patch
$i$ receives a cost $D_i^{(0)}$ that combines Euclidean distance with a
direct-path blockage penalty,
\begin{equation}
\label{eq:init_cost}
D_i^{(0)}
=
\frac{\bigl\|\mathbf{u}_{\text{tx}}-\mathbf{u}_i\bigr\|_2}
{\bigl(1-\beta(\mathbf{u}_{\text{tx}},\mathbf{u}_i)\bigr)^{\alpha_{\text{LoS}}}},
\end{equation}
where $\beta(\cdot,\cdot)\in[0,1]$ denotes the blockage between two
locations computed from $\mathcal{H}$, and $\alpha_{\text{LoS}}$
controls how strongly direct-path blockage inflates the cost. A patch
blocked along the direct path may still be reachable at lower total
cost through a less obstructed intermediate region. PILOT therefore
performs Dijkstra-style relaxation over the 8-connected neighborhood
$\mathcal{N}(i)$ of each settled patch $i$, updating each neighbor
$j\in\mathcal{N}(i)$ from step $t$ to $t+1$ as
\begin{equation}
\label{eq:relax_cost}
D_j^{(t+1)}
=
\min\!\left\{
D_j^{(t)},\;
D_i^{(t)}
+
\frac{\bigl\|\mathbf{u}_i-\mathbf{u}_j\bigr\|_2}
{\bigl(1-\beta(\mathbf{u}_i,\mathbf{u}_j)\bigr)^{\alpha_{\text{NLoS}}}}
\right\},
\end{equation}
where $\alpha_{\text{NLoS}}$ governs the blockage sensitivity of local
hops. The full procedure is summarized in Algorithm~\ref{alg:wavefront}, and example
propagation paths are shown in Fig.~\ref{fig:wavefront_context}(a).

Let $D_i$ denote the final relaxed cost of patch $i$ at convergence.
The deployed wavefront order is then defined as
\begin{equation}
\label{eq:pi_wavefront}
\pi_{\text{wavefront}}
=
\operatorname{argsort}_{i\in\mathcal{P}}\, D_i,
\end{equation}
where $\mathcal{P}=\{1,\dots,N\}$ indexes the $N$ patches and
$\pi_{\text{wavefront}}(n)$ returns the index of the patch with the
$n$-th smallest relaxed cost. The relaxation runs in $O(N\log N)$ on
the 8-connected patch graph with $N=256$, so the order is computed
once per scene at negligible cost relative to inference.

\begin{algorithm}[t]
\caption{Wavefront Propagation Order}
\label{alg:wavefront}
\begin{algorithmic}[1]
\REQUIRE Patch index set $\mathcal{P} = \{1,\dots,N\}$; transmitter
position $\mathbf{u}_{\mathrm{tx}}$; blockage function
$\beta(\cdot,\cdot)$ derived from the height map $\mathcal{H}$;
exponents $\alpha_{\mathrm{LoS}}$, $\alpha_{\mathrm{NLoS}}$.
\ENSURE Wavefront propagation order $\pi_{\text{wavefront}}$.
\FOR{each $i \in \mathcal{P}$}
    \STATE $D_i \leftarrow
           \|\mathbf{u}_{\mathrm{tx}} - \mathbf{u}_i\|_2
           \,\big/\,
           \bigl(1 - \beta(\mathbf{u}_{\mathrm{tx}}, \mathbf{u}_i)\bigr)
           ^{\alpha_{\mathrm{LoS}}}$
\ENDFOR
\STATE $\mathcal{V} \leftarrow \mathcal{P}$
\WHILE{$\mathcal{V} \neq \emptyset$}
    \STATE $i \leftarrow \arg\min_{k \in \mathcal{V}} D_k$
    \STATE $\mathcal{V} \leftarrow \mathcal{V} \setminus \{i\}$
    \FOR{each $j \in \mathcal{N}(i) \cap \mathcal{V}$}
        \STATE $w_{ij} \leftarrow
               \|\mathbf{u}_i - \mathbf{u}_j\|_2
               \,\big/\,
               \bigl(1 - \beta(\mathbf{u}_i, \mathbf{u}_j)\bigr)
               ^{\alpha_{\mathrm{NLoS}}}$
        \STATE $D_j \leftarrow \min\{D_j,\, D_i + w_{ij}\}$
    \ENDFOR
\ENDWHILE
\STATE $\pi_{\text{wavefront}} \leftarrow
       \operatorname{argsort}_{i \in \mathcal{P}}\, D_i$
\RETURN $\pi_{\text{wavefront}}$
\end{algorithmic}
\end{algorithm}

\subsubsection{Information-Entropy Analysis}
\label{subsubsec:information_entropy_analysis}

The wavefront order $\pi_{\text{wavefront}}$ is further analyzed
through an information-theoretic lens that links the choice of order
to the predictive uncertainty realized by a finite-capacity autoregressive model.
The analysis characterizes the prefix property induced by
$\pi_{\text{wavefront}}$.

Under the joint distribution $p(\mathbf{r}\,|\,\mathbf{e})$ defined in
Section~\ref{sec:formulation}, the chain rule of entropy gives
\begin{equation}
\label{eq:entropy_chain_rule}
\sum_{n=1}^{N}
H\bigl(r_{\pi(n)}\,|\,\mathbf{e}, r_{\pi(<n)}\bigr)
=
H\bigl(\mathbf{r}\,|\,\mathbf{e}\bigr),
\end{equation}
for every permutation $\pi$, so the aggregate conditional entropy is
invariant to ordering. Reordering cannot reduce the total uncertainty
when the model matches the true distribution.

This invariance does not constrain the predictive entropy of a
finite-capacity model with $p_\theta\neq p$. The predictive entropy of
$p_\theta$ at step $n$ under order $\pi$ is
\begin{equation}
\label{eq:predictive_entropy}
H_\pi^\theta(n)
=
-\sum_{c=1}^{|C|}
p_\theta\!\bigl(c\,|\,\mathbf{e}, r_{\pi(<n)}\bigr)
\log p_\theta\!\bigl(c\,|\,\mathbf{e}, r_{\pi(<n)}\bigr),
\end{equation}
which equals the entropy of the codebook-index softmax produced by
the decoder at step $n$. Different orders $\pi$ induce different
prefixes $r_{\pi(<n)}$, hence different per-step entropies
$H_\pi^\theta(n)$ and different averages
$\bar{H}^\theta(\pi)=\tfrac{1}{N}\sum_n H_\pi^\theta(n)$. Whether
$\pi$ helps the model depends on whether the prefix surfaces the
propagation-relevant predecessors of patch $\pi(n)$, a property
observed empirically in autoregressive language and image generation.

For the wavefront order, the predecessor containment property can be
stated explicitly. For each patch $i$, let
$\mathcal{P}^*(i)=\{j_0,j_1,\dots,j_{K-1}\}$ denote any
shortest-cost predecessor chain induced by the final cost graph from
$\mathbf{u}_{\text{tx}}$ to $i$. Along this chain, the recorded costs
satisfy
$D_{j_0}<D_{j_1}<\cdots<D_{j_{K-1}}<D_i$,
since every relaxation step adds a positive edge weight. When
$\pi_{\text{wavefront}}(n)=i$, sorting patches by ascending $D$ to
form $\pi_{\text{wavefront}}$ guarantees
\begin{equation}
\label{eq:predecessor_containment}
\mathcal{P}^*(i) \subseteq \bigl\{\pi_{\text{wavefront}}(1),\dots,
\pi_{\text{wavefront}}(n-1)\bigr\}.
\end{equation}
The wavefront prefix at the step that predicts patch $i$ thus contains
an entire shortest-cost predecessor chain to $i$. Under raster
ordering, \eqref{eq:predecessor_containment} does not hold in general,
since the predecessors on the propagation path to a shadowed patch may
lie outside the raster prefix block.

\begin{figure}[t]
  \centering
  \begin{minipage}[b]{0.48\columnwidth}
    \centering
    \includegraphics[width=\linewidth]{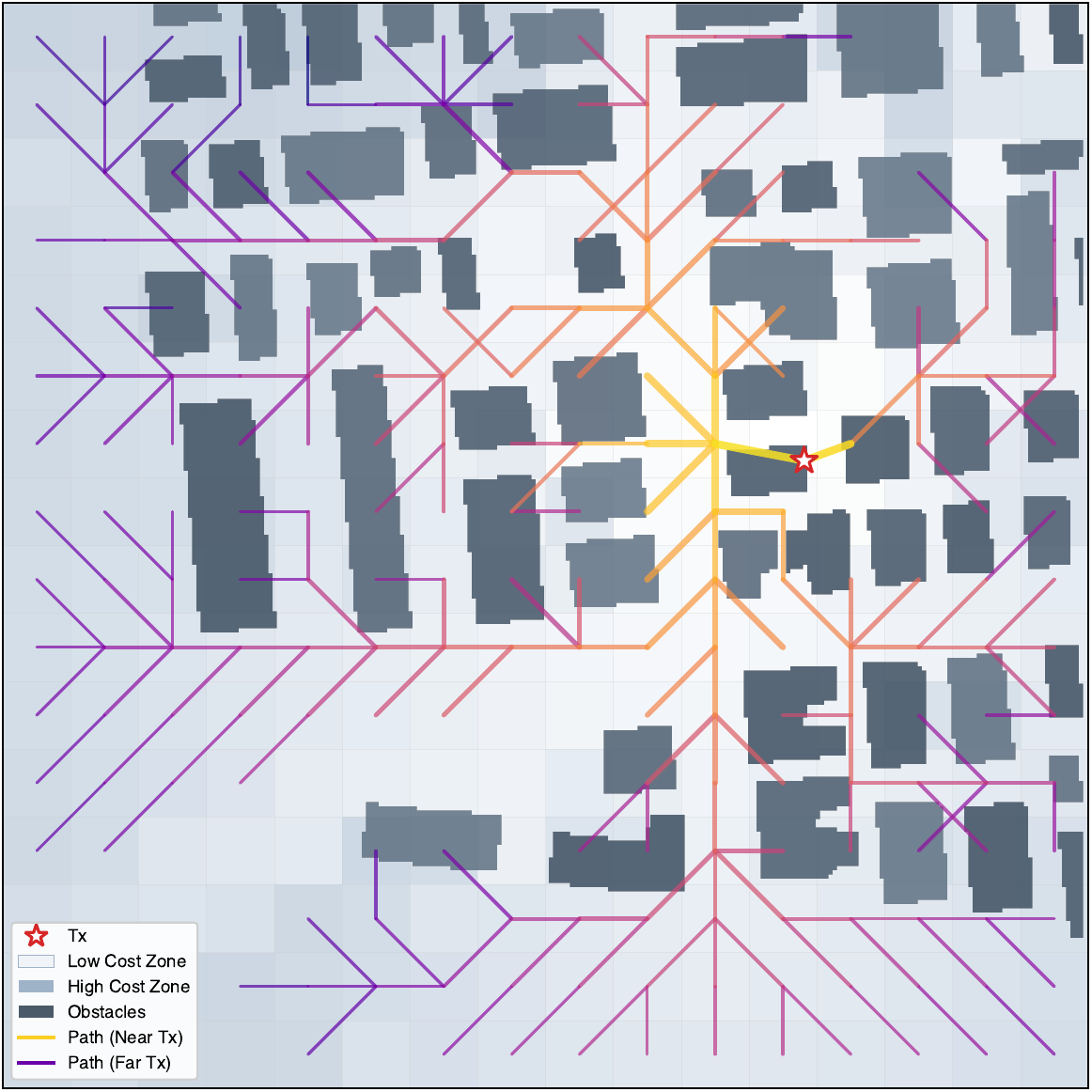}\\
    \vspace{0.5em}
    \small (a)
  \end{minipage}\hfill
  \begin{minipage}[b]{0.48\columnwidth}
    \centering
    \includegraphics[width=\linewidth]{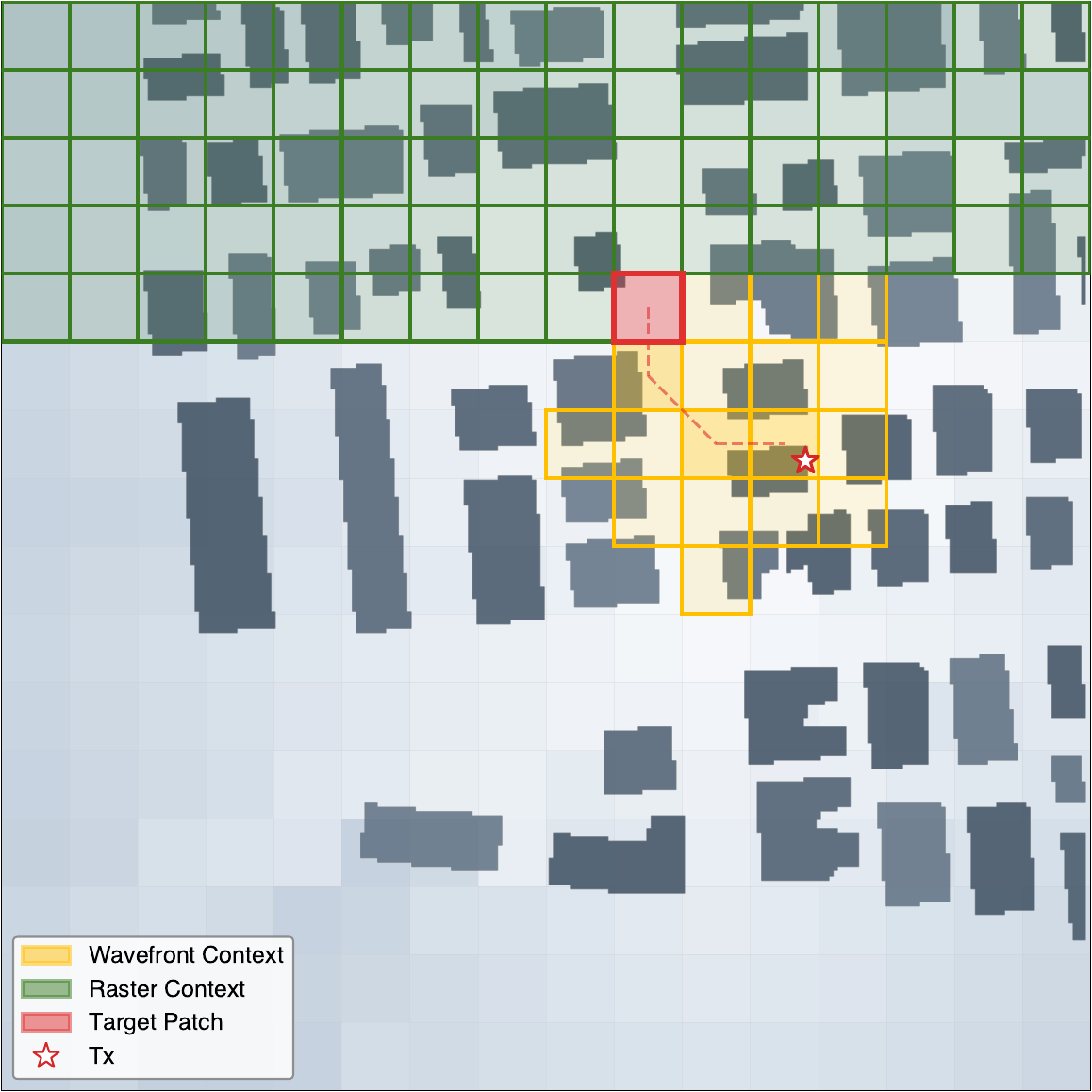}\\
    \vspace{0.5em}
    \small (b)
  \end{minipage}
  \caption{(a) Wavefront propagation paths in an urban scene, routing around blocked regions through lower-cost detours. (b) Visible prefix comparison for the same target patch (red): the raster-order prefix (green) consists of spatially preceding patches largely irrelevant to the target propagation state, whereas the wavefront-order prefix (yellow) concentrates along the propagation route to the target.}
  \label{fig:wavefront_context}
\end{figure}

By ranking patches according to $D$, the wavefront prefix exposes a
lower-cost surrogate propagation route that is more aligned with
blockage-aware reachability than a raster prefix, and this difference
is expected to manifest as a lower realized
$\bar{H}^\theta(\pi_{\text{wavefront}})$ on patches that depend on
long propagation paths. Fig.~\ref{fig:wavefront_context}(b) illustrates the prefix difference on
a representative target patch: under raster order the visible prefix
is dominated by spatially adjacent patches with no propagation link to
the target, while under wavefront order the visible prefix concentrates along
the lower-cost propagation route to the target. Section~\ref{subsubsec:conditional_entropy_verification} evaluates this prediction through inference-time
measurements of the realized predictive entropy.

\subsection{Environment-Aware Autoregressive Generation}
\label{subsec:environment_aware_ar_generation}

Wavefront ordering specifies which patch is predicted next, but not
what environment evidence should be exposed at that step. Using the
full environment sequence as a one-shot prefix would introduce many
regions unrelated to the current propagation state, dilute attention,
and increase decoding cost. PILOT instead uses one global CLS token
and step-aligned environment tokens. The CLS token provides
scene-level context, and the patch-aligned token $e_{\pi(n)}$ supplies
the environment cue for each target $r_{\pi(n)}$.

\subsubsection{Environment-Guided Context}
\label{subsubsec:environment_guided_context}

The environment input $\mathbf{E}\in\mathbb{R}^{256\times256\times3}$
is encoded by a ViT initialized from DINOv3 into $1+N=257$ tokens of
dimension 768, comprising one global CLS token and $N=256$
patch-aligned environment tokens $\{e_i\}$. An multilayer perceptron (MLP) projector then maps
each token to the decoder hidden dimension of 1024. The ViT and the
projector are fine-tuned end-to-end with the decoder.

The CLS token is prepended once, and the $N$ environment tokens are
interleaved with the radio tokens under the order $\pi$ to form the
input sequence
\begin{equation}
\label{eq:interleaved_sequence}
\bigl[\,
\text{CLS},
e_{\pi(1)},
r_{\pi(1)},
e_{\pi(2)},
r_{\pi(2)},
\dots,
e_{\pi(N)},
r_{\pi(N)}
\,\bigr],
\end{equation}
with causal masking applied throughout. At step $n$, the visible
prefix consists of the CLS token, the previously decoded
environment-radio pairs, and the current environment token
$e_{\pi(n)}$. Environment tokens enter the prefix incrementally under
the wavefront order, rather than as a fixed scene-wide prefix. The
corresponding $\pi$-conditioned likelihood factorizes as
\begin{equation}
\label{eq:pi_conditioned_likelihood}
\begin{aligned}
p_\theta\bigl(\mathbf{r}\,|\,\mathbf{e};\pi\bigr)
&= \prod_{n=1}^{N} p_\theta\bigl(r_{\pi(n)} \,\big|\, \text{CLS},
e_{\pi(1)}, r_{\pi(1)}, \\
&\hphantom{= \prod_{n=1}^{N} p_\theta\bigl(}
\dots, e_{\pi(n-1)}, r_{\pi(n-1)}, e_{\pi(n)}\bigr).
\end{aligned}
\end{equation}

\subsubsection{3D Spatial Registration}
\label{subsubsec:three_d_spatial_registration}

The ViT token carries environment semantics for the queried patch,
and 3D-RoPE assigns each matched pair $(e_i, r_i)$ to the same 3D
index $(x_i, y_i, z_i)$. The 1D rotary position embedding is defined
as
\begin{equation}
\label{eq:rope_1d}
\mathrm{RoPE}\bigl(\mathbf{q},m\bigr)
=
\left(
\bigoplus_{j=1}^{d/2}
\begin{bmatrix}
\cos(m\phi_j) & -\sin(m\phi_j) \\
\sin(m\phi_j) & \cos(m\phi_j)
\end{bmatrix}
\right)\mathbf{q},
\end{equation}
where $\bigoplus$ denotes the block-diagonal direct sum, $d$ is the
per-head rotary dimension, $m$ is the 1D position index, and
$\phi_j=10000^{-2(j-1)/d}$ is the $j$-th rotary frequency. For 3D
position indexing, each per-head query vector is partitioned into
three axis-specific subvectors as
\begin{equation}
\label{eq:q_split}
\mathbf{q}
=
\begin{bmatrix}
\mathbf{q}^{(x)} \\
\mathbf{q}^{(y)} \\
\mathbf{q}^{(z)}
\end{bmatrix},
\end{equation}
where the per-head rotary dimension is split approximately evenly
across the three axes. The 3D rotary embedding then applies the 1D
rotation to each subvector with its corresponding axis index,
\begin{equation}
\label{eq:rope_3d}
\mathrm{3D\text{-}RoPE}\bigl(\mathbf{q},x,y,z\bigr)
=
\begin{bmatrix}
\mathrm{RoPE}\bigl(\mathbf{q}^{(x)},x\bigr) \\
\mathrm{RoPE}\bigl(\mathbf{q}^{(y)},y\bigr) \\
\mathrm{RoPE}\bigl(\mathbf{q}^{(z)},z\bigr)
\end{bmatrix}.
\end{equation}
The same coordinate rule applies in 2D and 3D, with $z_i$
instantiated as the fixed receiver height in the 2D case and as the
per-slice height across the $N_z$ channels in the 3D case. The
decoder weights are shared across both settings.

\subsection{Training and Inference}
\label{subsec:training_and_inference}

During teacher forcing, the cross-entropy loss is evaluated only at
the steps whose target is a radio token. For each training
sample, the order is sampled uniformly from three candidates:
$\pi_{\text{wavefront}}$ from \eqref{eq:pi_wavefront};
$\pi_{\text{priorPL}}$, which ranks patches by ascending value of the
anchor map $\mathcal{M}_{\text{anc}}$ derived from frequency-aware
free-space and blockage information; and $\pi_{\text{truePL}}$, which
ranks patches by ascending value of the ground-truth radio map
$\mathbf{R}$ and serves as an oracle reference.

At inference, $\pi_{\text{wavefront}}$ is computed from $\mathbf{E}$
via Algorithm~\ref{alg:wavefront} and held fixed throughout decoding. The decoder
generates the $N=256$ radio tokens autoregressively by greedy
decoding, and the frozen tokenizer reconstructs
$\hat{\mathbf{R}}\in\mathbb{R}^{256\times256\times N_z}$ from the
predicted token grid. The number of autoregressive steps is fixed at
256 for both 2D and 3D settings.

\section{Evaluation}
\label{sec:evaluation}

\subsection{Experimental Setup}
\label{sec:setup}

We evaluate the proposed method on three tasks. Standard 2D radio map
construction measures in-domain reconstruction accuracy on held-out samples.
Zero-shot cross-domain transfer evaluates robustness to simultaneous shifts in
carrier frequency, transmitter geometry, and pathloss range without retraining.
3D radio map generation evaluates volumetric pathloss prediction
across receiver heights.

Table~\ref{tab:dataset} summarizes the datasets.
RadioMapSeer~\cite{levie2021radiounet} contains 56{,}000 ray-traced $256
\times 256$ pathloss maps at 5.9~GHz, with both transmitter and receiver fixed
at 1.5~m and building height fixed at 25~m.
RadioMap3DSeer~\cite{levie2021radiounet} uses the same horizontal resolution but
operates at 3.5~GHz, places the transmitter 3~m above the rooftop, and has a
narrower pathloss range, making it the target domain for zero-shot transfer.
UrbanRadio3D~\cite{wang2025radiodiff3d} provides a $256 \times 256 \times 20$
volumetric benchmark at 5.9~GHz with 1~m spatial resolution, realistic building
heights from 6.6 to 19.8~m, and receiver heights from 1 to 20~m. For pathloss
prediction, we use its 2.84 million pathloss maps.

We compare with five baselines from three paradigms:
\begin{itemize}
    \item \textbf{RadioUNet}~\cite{levie2021radiounet}: a cascaded two-UNet architecture that predicts the radio map in a single forward pass from environment-conditioned inputs and serves as the standard discriminative baseline for 2D construction.
    \item \textbf{RadioMamba}~\cite{jia2025radiomamba}: a hybrid Mamba-U-Net model that combines convolutional feature extraction with linear-complexity global context modeling and serves as a stronger single-pass discriminative baseline.
    \item \textbf{RME-GAN}~\cite{zhang2023rme}: a conditional adversarial reconstruction framework originally designed for sparse-measurement-conditioned radio map estimation. In this work, it is evaluated as an adapted variant without sparse measurements so that its conditioning setting matches the common 2D baseline setting.
    \item \textbf{RadioDiff}~\cite{wang2024radiodiff}: a conditional diffusion model for sampling-free 2D radio map construction that performs iterative denoising in a VAE latent space.
    \item \textbf{RadioDiff-3D}~\cite{wang2025radiodiff3d}: a 3D diffusion baseline with volumetric convolutional operators for pathloss generation on UrbanRadio3D.
\end{itemize}

RadioMapSeer is split into training, validation, and test sets of 500, 100, and
100 maps, respectively. For the 3D task, we use the 1--4~m subset of
UrbanRadio3D, which is divided at the file level into 90\% training files and
10\% test files, to match the evaluation setting of RadioDiff-3D. For the 2D
tasks, all models receive the building height map, transmitter position, and
pathloss anchor map as inputs. Since the original RME-GAN conditions on sparse
measurements, we remove its sparse-measurement branch and evaluate an adapted
version under the same 2D input setting. For the 3D task, the pathloss anchor
map is withheld, and each model receives only the volumetric environment
representation and transmitter encoding. Pathloss maps are min-max normalized to
$[0,1]$ over $[-47,-169]$~dB for cross-dataset comparability. NMSE, structural
similarity index measure (SSIM), and peak signal-to-noise ratio (PSNR) are
computed in the normalized domain, while root mean squared error (RMSE) is
reported in dB. All reproducible baselines are retrained with AdamW, a learning
rate of $10^{-4}$, batch size 32, and mixed-precision bf16 on two NVIDIA
RTX~4090 GPUs. Inference is measured on a single RTX~4090.

\begin{table}[htbp]
\caption{Dataset specifications}
\label{tab:dataset}
\centering
\resizebox{\columnwidth}{!}{%
\begin{tabular}{c|ccc}
\hline
Parameter & RadioMapSeer~\cite{levie2021radiounet} & RadioMap3DSeer~\cite{levie2021radiounet} & UrbanRadio3D~\cite{wang2025radiodiff3d} \\
\hline
Task type & Standard 2D estimation & Zero-shot generalization & 3D estimation \\
Dataset size & 56k & 56k & 2.8M$^{*}$ \\
Map size & $256 \times 256$ & $256 \times 256$ & $256 \times 256 \times 20$ \\
Pixel length & 1 m & 1 m & 1 m \\
Center carrier & 5.9 GHz & 3.5 GHz & 5.9 GHz \\
Building height & 25 m & 6.6--19.8 m & 6.6--19.8 m \\
Tx height & 1.5 m & 3 m above rooftop & 1.5 m \\
Rx height & 1.5 m & 1.5 m & 1.0--20.0 m \\
Pathloss range & $-47$ to $-147$ dB & $-75$ to $-111$ dB & $-92$ to $-169$ dB \\
\hline
\multicolumn{4}{l}{\footnotesize $^{*}$Reflects pathloss maps exclusively.} \\
\end{tabular}%
}
\end{table}

\subsection{Standard 2D Comparison}
\label{sec:2d}

\begin{figure*}[!htbp]
  \centering
  \includegraphics[width=\textwidth]{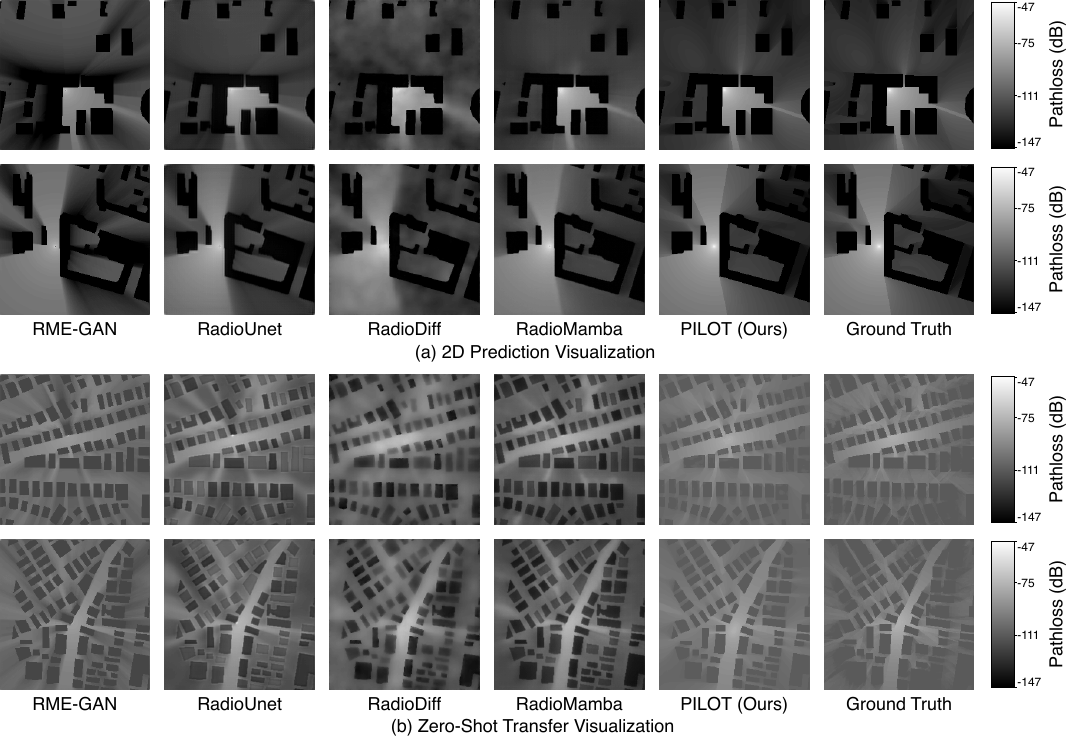}
  \caption{Predicted radio maps. (a) 2D construction on RadioMapSeer. (b) Zero-shot transfer to RadioMap3DSeer under frequency and geometry shift.}
  \label{fig:seer}
\end{figure*}

On RadioMapSeer, PILOT obtains the lowest NMSE as reported in
Table~\ref{tab:2d}. Relative to RadioMamba, NMSE drops from 0.0349 to 0.0316.
Qualitatively, PILOT better resolves shadow boundaries and diffraction
corridors as shown in Fig.~\ref{fig:seer}(a).

\begin{table}[htbp]
\caption{2D radio map construction on RadioMapSeer}
\label{tab:2d}
\centering
\resizebox{\columnwidth}{!}{%
\begin{tabular}{cccccc}
\hline
Model & NMSE $\downarrow$ & RMSE (dB) $\downarrow$ & SSIM $\uparrow$ & PSNR $\uparrow$ & Infer Time (s) $\downarrow$ \\
\hline
RME-GAN~\cite{zhang2023rme} & 0.1911 & 7.730 & 0.8707 & 22.40 & 0.0025 \\
RadioUnet~\cite{levie2021radiounet} & 0.1041 & 5.636 & 0.8991 & 25.25 & 0.0024 \\
RadioDiff~\cite{wang2024radiodiff} & 0.1111 & 5.785 & 0.9059 & 24.99 & 0.2960 \\
RadioMamba~\cite{jia2025radiomamba} & 0.0349 & 3.286 & 0.9322 & 29.93 & 0.0393 \\
\textbf{PILOT (Ours)} & \textbf{0.0316} & \textbf{3.076} & \textbf{0.9462} & \textbf{30.58} & 0.0252 \\
\hline
\end{tabular}%
}
\end{table}

\subsection{Zero-Shot Cross-Domain Transfer}

\begin{figure}[htbp]
  \centering
  \includegraphics[width=\columnwidth]{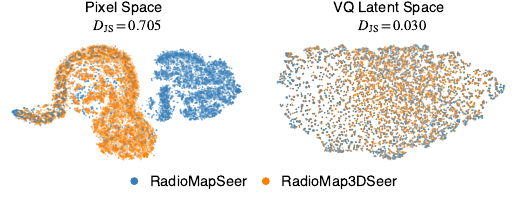}
  \caption{t-SNE of RadioMapSeer and RadioMap3DSeer distributions in pixel space and VQ latent space.}
  \label{fig:tsne}
\end{figure}

Table~\ref{tab:zeroshot} evaluates zero-shot transfer from RadioMapSeer to
RadioMap3DSeer under joint frequency, pathloss-range and geometry shift.
PILOT reduces NMSE by approximately 7.9\,\% over the strongest zero-shot
baseline RME-GAN, while RadioMamba degrades under domain
shift. Fig.~\ref{fig:seer}(b) shows that PILOT and RME-GAN output pathloss
ranges that match the target domain, whereas the other baselines remain
locked to the training range.
The shared pathloss anchor map anchors the pathloss range across frequencies,
and the VQ latent space reduces the Jensen--Shannon divergence
$D_{\mathrm{JS}}$ from 0.705 in pixel space to 0.030, as shown in
Fig.~\ref{fig:tsne} and Table~\ref{tab:vq_stats_comparison}.

\begin{table}[!t]
\centering
\caption{Distributional statistics of pixel-space and VQ-space representations across RadioMapSeer and RadioMap3DSeer}
\label{tab:vq_stats_comparison}
\renewcommand{\arraystretch}{1.2}
\resizebox{\columnwidth}{!}{%
\begin{tabular}{lcccccc}
\toprule
\multirow{2}{*}{\textbf{Representation}} & \multicolumn{2}{c}{\textbf{Norm. Entropy} $\uparrow$} & \multicolumn{2}{c}{\textbf{Gini Coeff.} $\downarrow$} & \multirow{2}{*}{$\boldsymbol{D_{\mathrm{JS}}}$ $\downarrow$} & \multirow{2}{*}{$\boldsymbol{\rho}$ $\uparrow$} \\
\cmidrule(lr){2-3} \cmidrule(lr){4-5}
 & \textbf{Seer} & \textbf{3DSeer} & \textbf{Seer} & \textbf{3DSeer} & & \\
\midrule
Pixel Space & 0.658 & 0.519 & 0.859 & 0.941 & 0.705 & 0.577 \\
\textbf{VQ Space} & \textbf{0.968} & \textbf{0.973} & \textbf{0.390} & \textbf{0.365} & \textbf{0.030} & \textbf{0.824} \\
\bottomrule
\end{tabular}
}
\end{table}

\begin{table}[htbp]
\caption{Zero-shot generalization comparison on RadioMap3DSeer}
\label{tab:zeroshot}
\centering
\resizebox{\columnwidth}{!}{%
\begin{tabular}{ccccc}
\hline
Model & NMSE $\downarrow$ & RMSE (dB) $\downarrow$ & SSIM $\uparrow$ & PSNR $\uparrow$ \\
\hline
RME-GAN~\cite{zhang2023rme} & 0.3115 & 3.873 & 0.7035 & 19.43 \\
RadioUnet~\cite{levie2021radiounet} & 0.5844 & 5.116 & 0.6699 & 17.18 \\
RadioDiff~\cite{wang2024radiodiff} & 0.5081 & 4.871 & 0.6327 & 17.50 \\
RadioMamba~\cite{jia2025radiomamba} & 0.6188 & 5.277 & 0.6255 & 16.92 \\
\textbf{PILOT (Ours)} & \textbf{0.2869} & \textbf{3.582} & \textbf{0.7451} & \textbf{20.24} \\
\hline
\end{tabular}%
}
\end{table}

\subsection{3D Radio Map Generation}

\begin{figure}[htbp]
  \centering
  \includegraphics[width=\columnwidth]{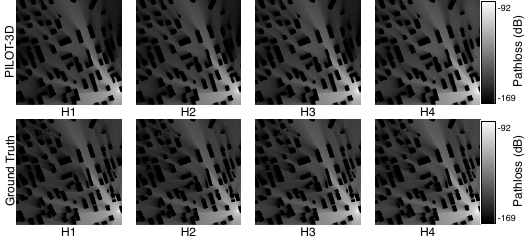}
  \caption{3D radio maps on UrbanRadio3D at receiver heights 1--4\,m.}
  \label{fig:3d_qual}
\end{figure}

Table~\ref{tab:3d} and Fig.~\ref{fig:3d_qual} report 3D results on
UrbanRadio3D at receiver heights 1--4\,m. PILOT in volumetric mode generates
multi-layer radio map tokens jointly and reduces NMSE from 0.0534 to 0.0120
relative to the 1000-step RadioDiff-3D diffusion baseline, at 0.049\,s per map
versus 121.76\,s. PILOT in height-conditioned mode produces a single-height map
by setting the $z$ coordinate of the 3D-RoPE embedding. The 200-step
RadioDiff-3D variant with 10\,\% sparse measurements still falls behind both
PILOT modes on NMSE and SSIM. Fig.~\ref{fig:entropy_curves}(c) shows that the 3D gradient loss
preserves inter-slice continuity across the four height levels, with the
90th-percentile vertical error reduced by 50\,\%.

\begin{table}[htbp]
\caption{3D radio map construction on UrbanRadio3D}
\label{tab:3d}
\centering
\resizebox{\columnwidth}{!}{%
\begin{tabular}{cccccc}
\hline
Model & NMSE $\downarrow$ & RMSE (dB) $\downarrow$ & SSIM $\uparrow$ & PSNR $\uparrow$ & Infer Time (s) $\downarrow$ \\
\hline
\begin{tabular}{@{}c@{}}RadioDiff-3D~\cite{wang2025radiodiff3d}\\($T = 200$)\end{tabular} & 0.3472 & 10.20 & 0.6453 & 19.57 & 24.3526 \\
\begin{tabular}{@{}c@{}}RadioDiff-3D\\($T = 1000$)\end{tabular} & 0.0534 & 5.03 & 0.8309 & 24.00 & 121.7630 \\
\begin{tabular}{@{}c@{}}RadioDiff-3D\\($T = 200+10\%$Samp.)\end{tabular} & 0.0550 & 3.70 & 0.8187 & 29.23 & 24.3526 \\
\textbf{PILOT volumetric} & 0.0120 & 2.60 & 0.9290 & 29.92 & \textbf{0.0486} \\
\textbf{PILOT height-cond.} & \textbf{0.0071} & \textbf{2.00} & \textbf{0.9498} & \textbf{32.49} & 0.1010 \\
\hline
\end{tabular}%
}
\end{table}

\subsection{Propagation Mechanism Analysis}
\label{sec:mechanism}

Section~\ref{sec:2d} established end-task accuracy.
This subsection investigates why propagation-aligned ordering improves generation by analyzing training convergence, comparing sequence variants at inference, and measuring predictive entropy per generation step.

\subsubsection{Generation Order Comparison}

\begin{figure*}[htbp]
  \centering
  \makebox[\textwidth][c]{%
    \begin{minipage}[b]{2.32in}
      \centering
      \includegraphics[width=\linewidth,height=1.85in]{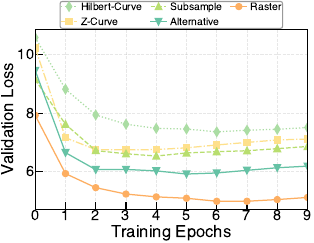}\\
      \vspace{0.5em}
      \small (a)
    \end{minipage}\hspace{0.1in}%
    \begin{minipage}[b]{2.32in}
      \centering
      \includegraphics[width=\linewidth,height=1.85in]{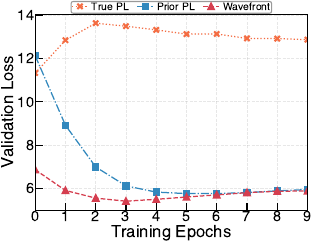}\\
      \vspace{0.5em}
      \small (b)
    \end{minipage}\hspace{0.1in}%
    \begin{minipage}[b]{2.32in}
      \centering
      \includegraphics[width=\linewidth,height=1.85in]{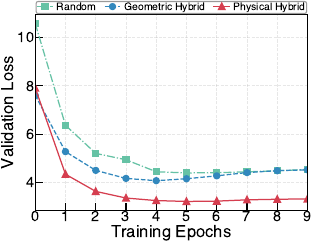}\\
      \vspace{0.5em}
      \small (c)
    \end{minipage}%
  }
  \caption{Validation cross-entropy by generation order. (a)~Geometric orders. (b)~Physics-guided orders. (c)~Hybrid strategies.}
  \label{fig:convergence}
\end{figure*}

\textit{Training stage.}
Fig.~\ref{fig:convergence} plots validation cross-entropy for eight individual orders and three hybrid strategies.
Among the geometric orders, raster scanning reaches the lowest loss at 5.12, while Hilbert, Z-curve, subsample, and alternative patterns settle between 6.19 and 7.51.
The physics-guided sequences, wavefront and prior-PL, converge to 5.89 and 5.95.
The true-PL order stalls at 12.87 because the ground-truth pathloss ranking varies abruptly across samples and provides no consistent spatial structure for the network to exploit.

The physical hybrid strategy reaches 3.33, well below the geometric hybrid at 4.53 and random ordering at 4.54.
Its constituent orders preserve physical structure through blockage-weighted costs and pathloss rankings, so each permutation supplies spatial context correlated with the propagation state.
Geometric mixtures, by contrast, merely rearrange scan patterns without physical alignment.
Mixing a bounded set of physics-guided orders also provides the sequence diversity needed for regularization while avoiding the intractable $N!$ permutation space, where model capacity would be diluted across uninformative arbitrary sequences.

\textit{Inference stage.}
Tables~\ref{tab:order_ablation_2d} and~\ref{tab:order_ablation_3d} evaluate a single physical hybrid checkpoint across eight generation orders.
On the in-domain RadioMapSeer benchmark in Table~\ref{tab:order_ablation_2d}, all non-oracle orders achieve NMSE between 0.0313 and 0.0326, indicating that inference-time ordering has minor effect when the test distribution matches the training data.
The true-PL oracle reaches 0.0255.
Although true-PL diverges during single-order training, the physical hybrid checkpoint encodes the correct mapping from propagation structure to tokens, and the oracle order then provides the sequence of lowest uncertainty at test time.

Under zero-shot transfer to RadioMap3DSeer in Table~\ref{tab:order_ablation_3d}, the gap between sequences widens.
The five geometric orders~\cite{pang2025randar} cluster near NMSE $= 0.2900$ because scene-agnostic scan patterns cannot adapt to unseen layouts and frequencies.
Both wavefront and true-PL achieve lower errors at 0.2869 and 0.2871.
The blockage-weighted Dijkstra cost thus serves as an accurate, training-free surrogate for the oracle order on unseen spatial distributions.

\begin{table}[htbp]
\caption{Ablation of generation orders on RadioMapSeer}
\label{tab:order_ablation_2d}
\centering
\resizebox{\columnwidth}{!}{%
\begin{tabular}{ccccc}
\hline
Order & NMSE $\downarrow$ & RMSE (dB) $\downarrow$ & SSIM $\uparrow$ & PSNR $\uparrow$ \\
\hline
Hilbert     & 0.0326 & 3.136 & 0.9448 & 30.42 \\
Z-curve     & 0.0316 & 3.091 & 0.9455 & 30.54 \\
Prior PL          & 0.0315 & 3.089 & 0.9455 & 30.54 \\
Subsample   & 0.0315 & 3.088 & 0.9455 & 30.55 \\
Raster      & 0.0314 & 3.085 & 0.9456 & 30.56 \\
Alternative & 0.0313 & 3.079 & 0.9457 & 30.57 \\
Wavefront        & 0.0316 & 3.076 & 0.9462 & 30.58 \\
True PL          & \textbf{0.0255} & \textbf{2.799} & \textbf{0.9492} & \textbf{31.39} \\
\hline
\end{tabular}%
}
\end{table}

\begin{table}[htbp]
\caption{Ablation of generation orders on RadioMap3DSeer}
\label{tab:order_ablation_3d}
\centering
\resizebox{\columnwidth}{!}{%
\begin{tabular}{ccccc}
\hline
Order & NMSE $\downarrow$ & RMSE (dB) $\downarrow$ & SSIM $\uparrow$ & PSNR $\uparrow$ \\
\hline
Subsample & 0.2899 & 3.601 & 0.7447 & 20.19 \\
Raster & 0.2899 & 3.601 & 0.7447 & 20.19 \\
Hilbert & 0.2900 & 3.601 & 0.7448 & 20.19 \\
Alternative & 0.2900 & 3.601 & 0.7448 & 20.19 \\
Z-curve & 0.2899 & 3.600 & 0.7448 & 20.19 \\
Prior PL & 0.2878 & 3.586 & 0.7451 & 20.23 \\
Wavefront & \textbf{0.2869} & \textbf{3.582} & 0.7451 & \textbf{20.24} \\
True PL & 0.2871 & \textbf{3.582} & \textbf{0.7452} & \textbf{20.24} \\
\hline
\end{tabular}%
}
\end{table}

\begin{figure*}[htbp]
  \centering
  \makebox[\textwidth][c]{%
    \begin{minipage}[b]{2.32in}
      \centering
      \includegraphics[width=\linewidth,height=1.85in]{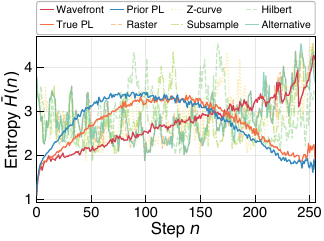}\\
      \vspace{0.5em}
      \small (a)
    \end{minipage}\hspace{0.1in}%
    \begin{minipage}[b]{2.32in}
      \centering
      \includegraphics[width=\linewidth,height=1.85in]{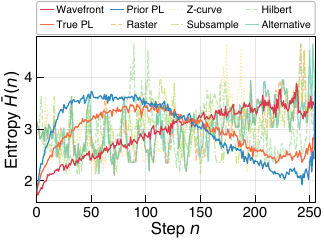}\\
      \vspace{0.5em}
      \small (b)
    \end{minipage}\hspace{0.1in}%
    \begin{minipage}[b]{2.32in}
      \centering
      \includegraphics[width=\linewidth,height=1.85in]{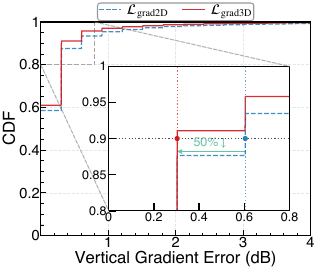}\\
      \vspace{0.5em}
      \small (c)
    \end{minipage}%
  }
  \caption{Predictive entropy and vertical gradient error measurements. (a)~Sample-averaged predictive entropy $\bar{H}(n)$ over 256 steps on RadioMapSeer. (b)~$\bar{H}(n)$ on RadioMap3DSeer under zero-shot transfer. (c)~CDF of vertical gradient error on UrbanRadio3D.}
  \label{fig:entropy_curves}
\end{figure*}
\subsubsection{Conditional Entropy Verification}
\label{subsubsec:conditional_entropy_verification}

We measure predictive entropy at each autoregressive step to quantify how the generation sequence affects model confidence.
At step $n$, the model outputs a logit vector $\mathbf{z}_n\in\mathbb{R}^{|\mathcal{C}|}$, and the corresponding entropy is
\begin{equation}\label{eq:entropy}
  H(n) = -\sum_{c=1}^{|\mathcal{C}|}
    \mathrm{softmax}(\mathbf{z}_n)_c \,\log\,
    \mathrm{softmax}(\mathbf{z}_n)_c.
\end{equation}
The physical hybrid checkpoint generates 256 tokens under each evaluated order.
We denote the sample-averaged predictive entropy at step $n$ as $\bar{H}(n)$.

Fig.~\ref{fig:entropy_curves}(a) shows $\bar{H}(n)$ on RadioMapSeer.
The true-PL and prior-PL sequences follow concave profiles that peak near the midpoint, where complex multipath effects concentrate.
The wavefront order increases monotonically from $\bar{H}(0)\!=\!1.2$ to $\bar{H}(255)\!=\!4.2$, confirming that generation begins with low-uncertainty near-field patches and defers complex shadow regions to later steps.
The five geometric orders oscillate without a clear trend and share an identical mean entropy of $\bar{H}\!=\!2.87$; rearranging a purely spatial scan pattern does not alter the aggregate conditional uncertainty.
The wavefront order yields $\bar{H}\!=\!2.72$, a 5.3\,\% reduction in mean predictive uncertainty.

Fig.~\ref{fig:entropy_curves}(b) repeats this measurement on RadioMap3DSeer under zero-shot transfer.
The three curve profiles persist across the domain shift: the wavefront order averages $\bar{H}\!=\!2.96$, compared with 3.01 for the geometric cluster.

\begin{figure}[htbp]
  \centering
  \includegraphics[width=3.5in]{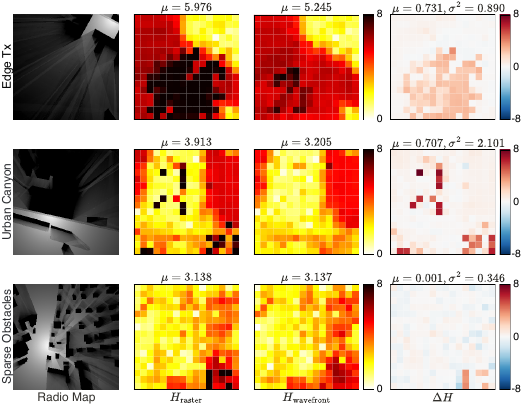}
  \caption{Spatial entropy maps for three propagation regimes.}
  \label{fig:entropy_heatmap}
\end{figure}

To localize this reduction geographically, we define $\Delta H = H_{\mathrm{raster}} - H_{\mathrm{wavefront}}$ per patch and evaluate three spatial regimes in Fig.~\ref{fig:entropy_heatmap}.
In the edge transmitter scenario, signals route around multiple obstacles before reaching deep non-line-of-sight regions, producing $\mu_{\Delta H}\!=\!0.731$ and $\sigma^2\!=\!0.890$; the positive $\Delta H$ values concentrate directly behind buildings.
In the urban canyon layout, the mean remains stable at $\mu_{\Delta H}\!=\!0.707$ but the variance rises to 2.101, with the uncertainty reduction localizing at canyon intersections and transition boundaries while straight corridors show negligible difference.
In the sparse obstacle environment, the metrics drop to $\mu_{\Delta H}\!=\!0.001$ and $\sigma^2\!=\!0.346$, confirming that the wavefront constraint introduces no detrimental bias when the geometry is structurally simple.

These results corroborate the formulation in Section~\ref{sec:formulation}:
propagation-aligned ordering reduces predictive entropy because each token
conditions on the correct physical history, and the magnitude of this reduction
scales with local propagation complexity.

\subsection{Ablation Studies}
\label{subsec:ablation_studies}
\subsubsection{Training Strategy}
Without pretraining, the autoregressive model reaches only NMSE 0.0828 on
RadioMapSeer. Fine-tuning from ImageNet-pretrained
LlamaGen-L~\cite{sun2024autoregressive} reduces this to 0.0316, a 62\,\%
drop. LoRA~\cite{hu2022lora} stops at 0.0559 because its low-rank subspace
cannot bridge the full domain gap from natural images to radio maps, and full
weight updates are needed as confirmed in Table~\ref{tab:training_strategy}.
Fig.~\ref{fig:entropy_curves}(c) shows that the 3D gradient loss
preserves inter-slice continuity, with the 90th-percentile vertical error
reduced by 50\,\%.

\begin{table}[htbp]
\caption{Ablation of training strategies on RadioMapSeer}
\label{tab:training_strategy}
\centering
\resizebox{\columnwidth}{!}{%
\begin{tabular}{ccccc}
\hline
Strategy & NMSE $\downarrow$ & RMSE (dB) $\downarrow$ & SSIM $\uparrow$ & PSNR $\uparrow$ \\
\hline
Training from Scratch & 0.0828 & 4.852 & 0.9063 & 26.57 \\
LoRA~\cite{hu2022lora} & 0.0559 & 4.123 & 0.9195 & 28.02 \\
Full Fine-tuning & \textbf{0.0316} & \textbf{3.076} & \textbf{0.9462} & \textbf{30.58} \\
\hline
\end{tabular}%
}
\end{table}

\subsubsection{Tokenizer Design}
The VQ-VAE tokenizer determines the upper bound of autoregressive generation,
because details lost during tokenization cannot be recovered later.
Table~\ref{tab:vq_ablation} quantifies the contribution of each component to the
tokenizer performance. The baseline VQ-VAE from
LlamaGen~\cite{sun2024autoregressive} suffers from severe codebook collapse,
with only 0.25\,\% utilization and poor reconstruction. Replacing the
convolutional projector with the linear projector of DiVeQ~\cite{vali2026diveq}
raises utilization to 99.7\,\% and reduces NMSE from 0.1507 to 0.0605.
Differentiable quantization~\cite{vali2026diveq} further reduces NMSE to 0.0253
by restoring gradient flow through the discrete bottleneck. Adding the
multi-scale gradient loss~\cite{Mathieu2016Deep} further improves boundary
fidelity and achieves the best results, with NMSE 0.0189, SSIM 0.9626, and PSNR
33.46\,dB.

\begin{table}[htbp]
    \centering
    \caption{Ablation of reconstruction components in the VQ-VAE tokenizer on RadioMapSeer}
    \label{tab:vq_ablation}
    \renewcommand{\arraystretch}{1.2}
    \resizebox{\columnwidth}{!}{%
        \begin{tabular}{cccccc}
            \toprule
            \textbf{Method} & \textbf{Codebook Util.} $\uparrow$ & \textbf{NMSE} $\downarrow$ & \textbf{RMSE (dB)} $\downarrow$ & \textbf{SSIM} $\uparrow$ & \textbf{PSNR} $\uparrow$ \\
            \midrule
            Baseline~\cite{sun2024autoregressive}     & 0.25\% & 0.1507 & 5.661 & 0.8365 & 25.15 \\
            + Linear Projector~\cite{zhu2025addressing}   & \textbf{99.7\%} & 0.0605 & 3.264 & 0.9213 & 30.11 \\
            + Diff. Quantization~\cite{vali2026diveq}   & 99.6\% & 0.0253 & 2.682 & 0.9534 & 32.10 \\
            + Gradient Loss~\cite{Mathieu2016Deep} & \textbf{99.7\%} & \textbf{0.0189} & \textbf{2.305} & \textbf{0.9626} & \textbf{33.46} \\
            \bottomrule
        \end{tabular}%
    }
\end{table}

\section{Conclusion}
\label{sec:conclusion}
This paper presented PILOT, a physics-integrated autoregressive framework for
unified 2D and 3D radio map construction. By coupling propagation-guided
generation with a frequency-aware pathloss anchor map, PILOT achieves strong
performance in standard 2D radio map estimation, enables zero-shot cross-domain
transfer without sparse measurements, and extends effectively to 3D radio map
generation. On the evaluated benchmarks, PILOT achieved up to $2500\times$
faster inference than the diffusion baseline while outperforming a 3D baseline
supported by 10\% measurements. These results show that physics-guided
autoregressive generation can produce accurate radio maps without sparse
measurements, at inference speeds suitable for real-time planning.

\bibliographystyle{IEEEtran}
\bibliography{refs}

\end{document}